\documentclass[a4paper,11pt]{article}
\usepackage{aaskaiid}
\usepackage{orcidlink}
\newcommand{\apj}{Astrophys. J.} % Astrophysical Journal
\newcommand{\apjs}{Astrophys. J. Suppl. Ser.} % Astrophysical Journal, Supplement
\newcommand{\aap}{Astron. Astrophys.} % Astronomy and Astrophysics
\newcommand{\jcap}{J. Cosmol. Astropart. Phys.} % Journal of Cosmology and Astroparticle Physics
\newcommand{\mnras}{Mon. Not. R. Astron. Soc.} % Monthly Notices of the RAS
\newcommand{\ssr}{Space Sci. Rev.} % Space Science Reviews
\title{An SKA-Low RM Grid for constraining the origin of cosmic magnetism}
\author[1]{Shane P. O'Sullivan\orcidlink{0000-0002-3968-3051}}
\ShortName{O'Sullivan et al.} % shortened name list fr header 
\author[2]{Franco Vazza}
\author[3]{Ettore Carretti\orcidlink{0000-0002-3973-8403}
}
\author[4]{Valentina Vacca\orcidlink{0000-0003-1997-0771}}
\author[4]{Francesca Loi}

\affiliation[1]{Departamento de Física de la Tierra y Astrofísica \& IPARCOS-UCM, Universidad Complutense de Madrid, 28040 Madrid, Spain}
\emailAdd{s.p.osullivan@ucm.es}
\affiliation[2]{Dipartimento di Fisica e Astronomia, Universitá di Bologna, via Gobetti 93/2, 40122 Bologna, Italy}
\affiliation[3]{INAF - Istituto di Radioastronomia, Via Gobetti 101, 40129, Bologna, Italy}
\affiliation[4]{INAF-Osservatorio Astronomico di Cagliari, Via della Scienza 5, I-09047 Selargius (CA), Italy}
\abstract{
Understanding the origin and evolution of cosmic magnetic fields is a key science goal for the SKAO. 
Recent advances in metre-wavelength (m-$\lambda$) Faraday rotation measure (RM) grids are enabling precision probes of cosmic magnetism, with implications extending to early-Universe physics, AGN feedback, and the magnetized circumgalactic medium.
Here we model the m-$\lambda$ polarized source counts to predict an RM Grid density with SKA-Low of  
$N(>P) \sim 5 ({P}/{100{\rm \mu Jy}})^{-0.75}\,\, {\rm deg}^{-2}  $, where $P$ is the polarized intensity detection threshold. 
This represents at least an order of magnitude improvement over the current state-of-the-art. 
For a representative wide-area SKA-Low AA4 survey covering 10,000 deg$^2$ in $\sim$3,200 hours, we predict more than 50,000 RMs. 
Coupled with an expected RM precision of $\sim$0.05 rad/m$^2$, SKA-Low promises to produce the leading RM Grid survey for constraining the origin of cosmic magnetism in the SKA era.\\ 
These predictions can be partially tested during the Science Verification phase using the AA* Sky Model data. For example, at a nominal detection threshold of 240~$\mu$Jy/beam (8 times the noise in Stokes $Q$ and $U$), we expect $\sim$2.6 RMs/deg$^2$ (5x the current best m-$\lambda$ RM Grid density). 
Combining both wide-area and all-sky data, SKA-Low could detect up to 100,000 m-$\lambda$ RMs across its observable sky.\\
Finally, we demonstrate new constraints on the origin of cosmic magnetism by comparing cosmological MHD simulations with the LOFAR m-$\lambda$ RMs, and highlight the transformative advances an SKA-Low RM Grid will enable for precision studies of cosmic magnetism.
}

\begin{document}
\maketitle

\section{Introduction}
Understanding the origin and evolution of cosmic magnetic fields is one of the key science goals for the SKAO \citep{heald_magnetism_2020,johnston-hollitt_using_2015}. Radio synchrotron radiation, its linear polarization, and the ensuing Faraday rotation provide some of the best ways of studying extragalactic magnetic fields \citep[e.g.][]{beckgaensler2004,vazza2015a,vanweeren2019,vernstrom_discovery_2021,vernstrom_s_2023,carretti_magnetic_2022-1}. 
In addition, there are many multi-wavelength tracers of the properties of cosmic magnetic fields, providing complementary constraints, such as $\gamma$-ray cascades \citep{neronov_evidence_2010}, ultra-high energy cosmic rays \citep{hackstein_simulations_2017}, the CMB \citep{planck_primordialB}, mm-polarized dust emission from galaxies \citep{geach_polarized_2023}, and the Lyman-$\alpha$ forest \citep{pavicevic_constraints_2025}. 

Recent studies using Faraday rotation measure (RM) Grids at metre-wavelengths \citep{van_eck_polarized_2018-3,riseley_polarised_2020,osullivan_faraday_2023} emphasize their importance to understanding the origin of cosmic magnetism \citep{carretti_nature_2024,neronov_revision_2024}, the physics of the early Universe \citep{neronov_nanograv_2021,mtchedlidze_intergalactic_2024}, the maximum extent of feedback from AGN and galaxy outflows \citep{blunier_constraint_2024,bondarenko_probing_2024,va25a}, and the properties of the circumgalactic medium \citep{heesenDetectionMagneticFields2023a,ramesh_azimuthal_2023-1}. While the focus of this chapter is on using an SKA-Low RM Grid to advance our understanding of the origin of cosmic magnetism, we note that the science applications are much broader than this. For example, RM Grids can be used to probe the inventory of matter within and beyond virialised halos out to cosmic noon \citep{anderson2024,bockmann_probing_2023,Malik2026}, which has strong synergies with other current and planned facilities, such as the Simons Observatory \citep{simons2019}, ELT-MOSAIC \citep{MOSAIC} and At-LAST \citep{atlast}. 

The current weight of evidence indicates the presence of relatively strong primordial magnetic fields, with $B_{\rm 0,Mpc}\sim0.4$~nG \citep[][and references therein]{cava25}. Such primordial fields can have a considerable impact on early galaxy evolution \citep{sanati_dwarf_2024}, in addition to potentially alleviating the Hubble tension \citep{jedamzik_relieving_2020}. 
Being able to robustly constrain the properties of primordial fields leads us into an era where we have a new observational probe of early-Universe physics \citep{brandenburg_magnetogenesis_2024}, with exciting links to the production of stochastic gravitational waves signatures \citep{neronov_nanograv_2021,agazie_nanograv_2023}.  

To predict the properties of an SKA-Low RM Grid, in Section~\ref{sec:model} we use models of extragalactic radio source counts and constrain them with our current best knowledge of the m-wavelength polarized sky. In Section~\ref{sec:lowRMgrid}, we use these constrained models to extrapolate the polarized-source counts to SKA-Low AA4 sensitivities in order to outline nominal SKA-Low RM Grid survey possibilities. We then highlight two complementary approaches to constraining the properties of primordial magnetic fields, using the redshift evolution of the RM (Section~\ref{sec:newconstraints}). 
%(Sections~\ref{sec:RRMz},\ref{sec:DRMdz}). 
Finally, we list some of the exciting opportunities for new discoveries provided by a wide-area SKA-Low RM Grid (Section~\ref{sec:advances}). 

%\begin{itemize}
%    \item cosmic B in general: gamma-ray (Neronov), radio synchrotron (Vernstrom), Faraday rotation (Beck \& Gaensler 2004), CMB (Planck), UHECR (Hackstein), lyman-alpha (recent ref), zeeman? 
%    \item RM Grids for cosmic magnetism in general
%    \item Why m-wavelength RM Grids are particularly powerful: they sample clean lines of sight due to depolarization effects, and their errors are much smaller. One could select a low-depolarization sub-sample from a cm-wavelength RM Grid, but the errors on individual data points are larger, so the final statistical precision depends on the relative number of sources. 
%    \item summary of recent advances using m-wavelength RM Grids (Carretti, Pomakov, Neronov etc.) for filaments, AGN feedback extent, primodial vs. astrophysical. 
%    \item complementary cm RM Grids: provide better opportunity for robust GRM models, and probe wider range of cosmic environments (cluster, groups, etc. studies) 
%    \item Focus of this chapter: what advances can be achieved by SKA-Low AA4? Need AA* numbers for talk.  Exploitation of various possible RM Grid survey(s) based on the extrapolation of our current best knowledge of the polarized sky at m-wavelengths.  
%\end{itemize}

%--------------------------------------------------------------------
\section{Polarized-source counts at SKA-Low frequencies}
\label{sec:model}

The Tiered Radio Extragalactic Continuum Simulation \citep[T-RECS;][]{bonaldi_tiered_2019,bonaldi_tiered_2023}, provides model catalogues of the radio-continuum sky from 150 MHz to 20 GHz. It simulates two primary populations: Active Galactic Nuclei (AGN) and Star-Forming Galaxies (SFG), along with their respective sub-populations. The outputs have been validated against up-to-date data compilations, showing excellent agreement in luminosity functions, number counts in total intensity, and clustering properties. The T-RECS code is publicly available and can be used to produce radio source catalogs with user-defined frequencies, areas, and depths. This makes it a valuable tool for extrapolating to what can be observed from surveys with the SKA-Mid and SKA-Low telescopes.

An updated 150 MHz model was developed by \citet{lin_new_2024} based on the LOFAR Deep Field Data Release 1 \citep{best2023}. This work recast radio source classifications into High Excitation Radio Galaxies (HERG), Low Excitation Radio Galaxies (LERG), Radio-Quiet AGN (RQ-AGN), and SFG. They found that the simulated source counts more closely match the observed data at $z>4$ than the original T-RECS, while also reproducing the observed differential source counts in total intensity at the faint end (0.1–1 mJy). They also provide updates to the public T-RECS code\footnote{\url{https://github.com/nxdtxdyka/f150\_ps\_simulation}}.  

T-RECS incorporates polarized emission across all frequency bands, characterized statistically for each population. However, while it produces robust results compared to 1.4 GHz polarized source counts, recent polarization observations near 150 MHz \citep{van_eck_polarized_2018-3,riseley_polarised_2020,osullivan_faraday_2023,piras_lofar_2024} show that it significantly over-predicts the polarized source counts at m-wavelengths. 
This is due to the effects of Faraday depolarization at m-wavelengths, which  suppresses the polarized emission for Faraday dispersions within the telescope beam of $\gtrsim 0.3$~rad/m$^2$ \citep{stuardi_lofar_2020,mahatma_low_2020}. Here we modify the updated T-RECS model output at 150 MHz to more accurately represent the polarized source counts at 150 MHz and their known statistical properties, with the aim of using it to extrapolate to SKA-Low sensitivities. 

\subsection{Observed properties of polarized sources at 150 MHz}
\label{sec:obs}
The largest catalogue of polarized sources at m-wavelengths to date is provided by the LOFAR Two-Metre Sky Survey (LoTSS) Data Release 2 (DR2) RM Grid \citep[][OS23]{osullivan_faraday_2023}, with 2,461 polarized sources  (with corresponding RMs) over a sky area of 5,720 deg$^2$. They used a polarized source detection threshold of 8 times the local noise in Stokes $Q$ and $U$, which translates to $\sim$0.5~mJy/beam (99.5\% of sources are detected above this threshold). The polarized sources were extracted from data at an angular resolution of 20$''$, across a frequency range of 120 to 168 MHz with a spectral resolution of 97.6~kHz. The median degree of polarization of these sources is 1.8\%, with $\sim$25\% of sources classified as compact/unresolved and $\sim$75\% as resolved (with a median Largest-Angular-Size of $\sim$$1'$ and a median linear size of $\sim$400~kpc). Host galaxy identifications were obtained for 88\% of sources, resulting in a redshift estimate for 79\% of the sources (37\% photometric, 42\% spectroscopic). The median spectral luminosity at 150 MHz is $5\times10^{26}$ W/Hz. 

Deeper polarization results at 150 MHz ($\sim$4 times deeper than OS23) were provided by \citet{piras_lofar_2024}, across 25 deg$^2$ of the ELAIS-N1 field, at a resolution of 6$''$. They found 31 polarized sources, with similar statistical properties to the LoTSS DR2 RM Grid. They also provide the first robust polarized source counts at 150 MHz, for flux densities $>0.2$~mJy (which we reproduce in Fig.~\ref{fig:counts}). 
For comparison purposes, we crudely estimate the polarized source counts for the LoTSS DR2 RM Grid, shown also in Fig.~\ref{fig:counts}. For this we assume a uniform polarized noise distribution across the entire DR2 area, which is inaccurate due to the presence of bright sources,  field-to-field quality variations (such as poor ionosphere conditions), and the absence of mosaicking in $Q$ and $U$. However, we find these estimates to be a useful lower bound for consistency checks at higher polarized flux densities and compared to the Piras et al.~results. It's also worth nothing the decreased effect of beam depolarization due to the $\sim$3 times higher angular resolution of the Piras et al.~observations, which could also help explain the differences. 

No polarized emission from any SFG has been found at 150 MHz (all identified extragalactic sources thus far are either radio galaxies or blazars). Therefore, in our modelling we only consider the radio AGN population from T-RECS. We consider this the conservative approach, since even with much deeper observations with SKA-Low, it may remain challenging to detect a statistical sample of SFGs at m-wavelengths due to the known significant depolarization effects \citep{stil_integrated_2008}. 

%Move/merge with other text: \textcolor{blue}{Thus, to more accurately model the polarized sources at m-wavelengths, we need to know the general polarized properties of the parent radio AGN (excluding the SFGs). Robustly separating radio AGN and SFGs is challenging, as it requires high-quality radio data in addition to substantial multi-wavelength data (e.g. Best+23, Hardcastle+25). Using the LoTSS DR2 data and results from the LOFAR Deep Fields, \citet{hardcastle2025} estimates the number density of radio AGN at flux densities $\gtrsim 1$~mJy as $\sim$100 to 170~deg$^{-2}$. Comparing to the LoTSS DR2 RM Grid detections, this implies a radio AGN detection fraction of $\gtrsim$0.3 to 0.5\% at 150 MHz. This can be considered a lower limit due to the conservative polarized source selection in \citet{osullivan2023}, where the focus was on reliability rather than completeness, in addition to the impact of depolarization from the residual ionosphere RM correction errors (which dominate the RM error budget). }

\subsection{Modelling the polarized source counts at 150 MHz}
\label{sec:polcounts}

\begin{figure}
        \centering
        \includegraphics[width=0.48\linewidth,clip=true,trim=0.5cm 0cm 0.5cm 0cm]{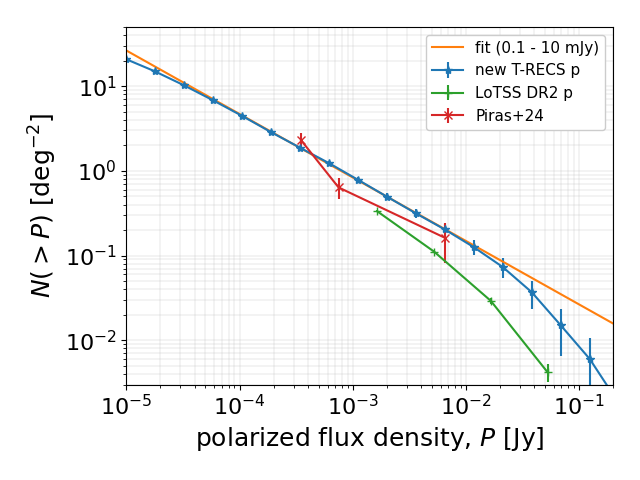}
        \includegraphics[width=0.48\linewidth,clip=true,trim=0.5cm 0cm 0.5cm 0cm]{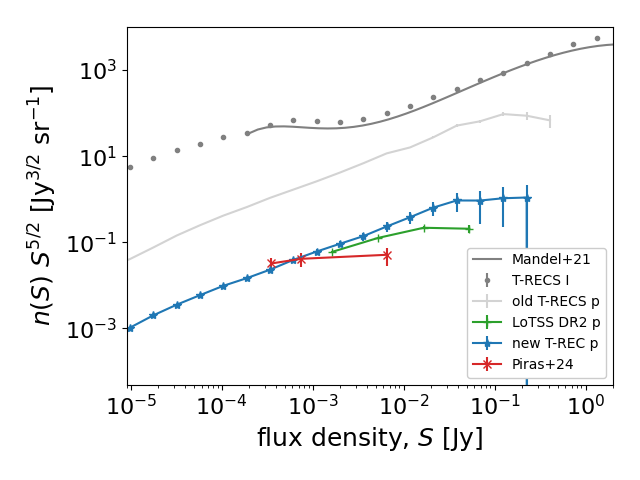}       
        \caption{
        Left: Cumulative polarized-source number counts: 
        updated T-RECS polarization (blue), 
        LOFAR deep field polarized source counts from \citet{piras_lofar_2024} (red), 
        LoTSS DR2 RM Grid counts (green), 
        power-law fit to model between 0.1 and 10 mJy (orange). 
        Right: Euclidean-normalised source counts:  
        T-RECS total intensity (grey points), LOFAR total intensity deep field counts from \citet{mandal_extremely_2021} (grey line), 
        previous T-RECS polarization (light grey), new polarized source counts (same as left plot). 
        %updated T-RECS polarization (blue), 
        %LOFAR deep field polarized source counts from \citet{piras_lofar_2024} (red), 
        %LoTSS DR2 RM Grid counts (green). 
        }
        \label{fig:counts}
\end{figure}

To model the polarized source counts we start from the T-RECS total intensity model for AGN (labelled as HERGs and LERGs in the updated 150 MHz version, with \textsc{RadioClass==1} and 2, respectively). We simulate a 200~deg$^{2}$ area to have a sufficiently large sample of bright radio galaxies ($\sim$50,000~$ > 1$~mJy) to compare statistically with the LoTSS DR2 AGN sample ($\sim$500,000~$ > 1$~mJy, in \citet{hardcastle2025} with $\sim$20\% having reliable size estimates). 
%In T-RECS-150 the number density of HERGs and LERGs greater than 0.1 mJy is $\sim$1000~deg$^{-2}$, while the number density above 1 mJy is $\sim$200~deg$^{-2}$. This is broadly consistent with LoTSS DR2 (for an optical ID fraction of 0.58) and the LOFAR Deep Fields estimate of $\sim$170~deg$^{-2}$. 
The median luminosities of the bright AGN are comparable ($10^{25}$~W/Hz), but the median angular size in T-RECS is $\sim$2.5 times smaller than in LoTSS. Thus, we increased all T-RECS sizes by a factor of 2.5.  

To model the polarized sources, we assign a degree of polarization ($p$) to each Stokes $I$ source randomly from a log-normal distribution with parameters $\mu=0.49$ and $\sigma=1.06$. These values were obtained by fitting to the observed $p$ distribution of the LoTSS DR2 RM Grid. %, Fig.~\ref{fig:fracpol}). 
We retain the drawn $p$ value if it is in the range 0.05 to 30\%, (as in the LoTSS DR2 RM Grid) and redraw otherwise. 
%We also imposed 
%a correlation between $p$ and linear size and 
%an anti-correlation of $-0.1$ between $p$ and redshift, to better match the known trend at 150 MHz \citep{carretti_magnetic_2022-2}. 
We split the sample into resolved (size $>$ 20$''$) and compact (sizes between 10$''$ and 20$''$, and linear size $> 20$~kpc, to avoid selecting low-luminosity AGN), and randomly retain only 25\% of the compact sample. This is done to match the statistics of the resolved and compact sources found in the LoTSS DR2 RM Grid. To reproduce the observed polarized source counts at the faint end (Fig.~\ref{fig:counts}), we randomly retained 35\% of all the sources assigned polarization.
%We also remove 50\% of the bright, resolved sources to account for surface brightness limitations on the detection of polarized emission. \textbf{what is the final fraction of resolved vs unresolved? currently 80/20\% so not far off}. 

This produced an output median $p$ of 1.9\%, for the sources assigned polarization (close to the observed median in the LoTSS DR2 RM Grid of 1.8\%). The median luminosity of the polarized sources remains comparable to the observations, while the boosted sizes closely match the size distribution of the observed polarized sources (all sources $<5$~Mpc and median of $\sim$300 kpc). Furthermore, the model matches the \citet{piras_lofar_2024} source counts at the faint end, while over-predicting the LoTSS DR2 counts at the bright end. We retain this over-prediction given that the LoTSS DR2 counts are approximate and likely significantly underestimated (Section~\ref{sec:obs}). 

For a power-law fit to the polarized source counts model for polarized flux densities ($P$) between 0.1 and 10 mJy, one obtains a polarized source number density ($N$) at 150 MHz of 
\begin{equation}
N(>P) \sim 5 \bigg(\frac{P}{100{\rm \mu Jy}} \bigg)^{-0.75}\,\, {\rm deg}^{-2}  , 
\label{eqn:pcounts}
\end{equation} 
providing a useful parameterisation for predictions within this range. 
It should be kept in mind that this model is mainly constrained at the faint end by the results from a single LoTSS Deep Field covering 25 deg$^2$ of sky at 6$''$. Thus, our predictions are subject to significant uncertainties, such as cosmic variance, in addition to the polarization performance of the LOFAR telescope. Variations up to a factor of 2 in the predicted number of polarized sources at a given detection threshold between 0.1 and 10 mJy should not be unexpected.

\section{SKA-Low RM Grid}
\label{sec:lowRMgrid}

To assess the potential scientific gains of nominal SKA-Low RM Grid surveys, we can use the 150 MHz polarized source counts model (Section~\ref{sec:model}) to extrapolate (with significant uncertainty) to SKA-Low-relevant flux-density detection thresholds. We consider two broad categories of survey below: a wide-area survey of $\sim$10,000 deg$^2$ of extragalactic sky to maximise the total number of RMs with redshifts, and an ultra-deep survey of $\sim$25 to 100 deg$^2$ to explore the discovery potential of future SKA-Low polarization surveys. 
%\textbf{FoV is about 25 deg$^2$; so 100 deg$^2$ considering 4 deep fields, either contiguous or not.}

Adopting ``Zoom Mode 4'' from the SKA-Low Sensitivity Calculator User Guide\footnote{\url{https://sensitivity-calculator.skao.int/low}}, we consider a frequency resolution of $\sim$100~kHz and a bandwidth of $\sim$200~MHz (for context, the LoTSS RM Grid used 97.6~kHz channels and 48 MHz of bandwidth). The optimal frequency range is from 100 to 300 MHz to balance detectability of polarized emission with RM precision.\footnote{Although the RM precision increases with $\lambda^2$ coverage, only one polarized radio galaxy has been detected below 100 MHz \citep{osullivan_faraday_2018-3}, mostly likely due to depolarization effects. } 
A central frequency of 200 MHz should lead to a slightly higher number density of polarized sources compared to 150 MHz (due to the reduced impact of wavelength-dependent depolarization); although the sensitivity-weighted central frequency may be closer to 150 MHz, depending on radio-frequency interference (RFI) and the final performance of the array. We assume that the residual level of wide-field instrumental polarization across the frequency band can be calibrated to 1\% of Stokes $I$ or better \citep[e.g.][]{VanWeeren2026}. The presence of RFI would also impact the quality of the Rotation Measure Spread Function (RMSF). However, as long as the lower half of the band remains largely unaffected, then the FWHM of the RMSF should be $\sim$0.5 rad/m$^2$ for this setup. 

\subsection{An SKA-Low wide-area RM Grid survey}
\label{sec:widearea}
Using the SKA-Low sensitivity calculator for AA4, an integration time of 8 hrs with a Briggs robust imaging weighting of $-1$ gives an rms sensitivity of $\sim$5~$\mu$Jy/beam at an angular resolution of $\sim$5$''$ (2 to 3 times above the confusion noise at this resolution). 
For the same input parameters, but with AA*, the sensitivity calculator gives an rms  of $\sim$7~$\mu$Jy/beam and a similar resolution. Considering the likely impact of RFI across the observing bandwidth and variable ionospheric conditions, in addition to the final array performance, we adopt an rms sensitivity of 10~$\mu$Jy/beam for a single 8 hr observation as a conservative estimate for a typical extragalactic field. 
%To be conservative, we consider a single 8 hr observation 
For a Field of View (FoV) of 5~deg, a 10,000 deg$^2$ survey could be completed in $\sim$3,200 hours of observing time. 

For a polarized source detection threshold of 80~$\mu$Jy (i.e.~$8\sigma_{QU}$), Eqn.~\ref{eqn:pcounts} predicts $\sim$5.9 RMs/deg$^2$, implying a total count of $\sim$59,000 RMs for a 10,000 deg$^2$ survey. This would be $\gtrsim20$ times the number of known RMs at m-wavelengths (OS23), providing a dramatic improvement in the discovery potential for cosmic magnetism science (assuming the majority will have host galaxy identifications and redshift estimates). 
The RM error at the detection threshold would be $\sim$0.03~rad/m$^2$.  However, the current best ionospheric RM modelling \citep{mevius2018} leaves residual errors of $\sim$0.05 rad/m$^2$ (OS23). Therefore, the expected level of the residual ionosphere RM errors means the RM precision would not improve for higher S/N sources (as it dominates even at the detection threshold). 

\textbf{Context within RM Grid landscape:} For comparison with other RM Grid survey configurations (i.e.~area vs.~depth vs.~frequency band), we define a figure-of-merit (FoM) related to the statistical precision of a sample of RMs: ${\rm FoM}=10^{-3}\sqrt{N_{\rm total}}/\delta{\rm RM}$, combining the RM number density and survey area ($N_{\rm total}$ is the total number of RMs), as well as the frequency coverage and survey depth ($\delta{\rm RM}$ is the RM error at the detection threshold). The 10$^{-3}$ factor normalises to numbers of order unity. The higher the FoM the better. The nominal FoM for the wide-area SKA-Low RM Grid ranges from ${\rm FoM_{SKA{\text -}Low,wide}}=5$ to 8, limited on the low end by the assumed residual ionosphere RM error, with the high end corresponding to the true RM precision at the detection threshold. 

\begin{table}[h]
	\centering
	\caption{Comparison of the Figure of Merit (${\rm FoM}=10^{-3}\sqrt{N_{\rm total}}/\delta{\rm RM}$) for various RM Grid surveys, combining RM number density, survey area, and $\lambda^2$-coverage (higher values of FoM are better). The resolution for SKA-Low and Mid telescopes are nominally for 150 MHz and 1 GHz, respectively. }
	\label{tab:FoM}
	\begin{tabular}{lcccccr} % four columns, alignment for each
		\hline
		Survey & Resolution & Sky area & $\delta{\rm RM}$ & Freq. & RMs/deg$^2$ & FoM \\
		       & [arcsec] & [deg$^2$] & [rad~m$^{-2}$] & [MHz] & [deg$^{-2}$] &   \\
		\hline
		SKA-Low wide-area & 5 & 10,000 & 0.05 & 100--300   & 5.9  & 5\\
		LoTSS DR2         & 20 & 5720   & 0.05 & 120--168   & 0.43 & 1\\
		SKA-Mid wide-area & 0.5 & 10,000 & 1.5  & 950--1760  & 100  & 0.7\\
		ASKAP-POSSUM      & 20 & 20,000 & 1.5 & 800--1100  & 40   & 0.6\\
		  NVSS              & 45 & 34,000 & 10   & $\sim$1400 & 1    & 0.02\\

        \hline
	\end{tabular}
\vspace{-0.5cm}
\end{table}
%\begin{table}[h]
%	\centering
%	\caption{Comparison of the Figure of Merit (${\rm FoM}=10^{-3}\sqrt{N_{\rm total}}/\delta{\rm RM}$) for various RM Grid surveys, combining RM number density, survey area, and $\lambda^2$-coverage (higher values of FoM are better).}
%	\label{tab:FoM}
%	\begin{tabular}{lccccr} % four columns, alignment for each
%		\hline
%		Survey & Sky area & $\delta{\rm RM}$ & Freq. & RMs/deg$^2$ & FoM \\
%		       & [deg$^2$] & [rad~m$^{-2}$] & [MHz] & [deg$^{-2}$] &   \\
%		\hline
%		SKA-Low wide-area & 10,000 & 0.05 & 100--300  & 5.7 & 5\\
%		LoTSS DR2         & 5720   & 0.05 & 120--168  & 0.43 & 1\\
%		ASKAP-POSSUM      & 20,000 & 3.35 & 800--1100 & 40  & 0.3\\
%		SKA-Mid wide-area & 10,000 & 3.3  & 950--1760 & 100 & 0.3\\
		%LoTSS DR2 & 5720 & 0.05 & 120--168 & 0.4 & 1\\

 %       \hline
%	\end{tabular}
%\end{table}

Even assuming the limiting scenario of no improvements in the current ionsphere RM correction methods, the ${\rm FoM_{SKA{\text -}Low,wide}}$ exceeds all other ongoing and planned RM surveys (Table~\ref{tab:FoM}). For example, the LoTSS RM Grid has ${\rm FoM_{LoTSS,DR2}}\sim1$, 
% assume 0.5 RMs/deg and 12k deg$^2$ area = 6k RMs, what I expect 
% note that (20k - 7k Gplane from +/-10 deg) = 13k deg$^2$
while ${\rm FoM_{POSSUM}}\sim0.6$ for ASKAP-POSSUM and 
${\rm FoM_{SKA{\text -}Mid,wide}}\sim0.7$ for SKA-Mid. 
For POSSUM we use a median $\delta{\rm RM_{med}}=1.5$~rad/m$^2$ for 800 to 1088 MHz, with 40 RMs/deg$^2$ and the full 20,000 deg$^2$ area \citep{vanderwoude_prototype_2024,gaensler_polarisation_2025}, while 
for SKA-Mid we use $\delta{\rm RM_{med}}=1.5$~rad/m$^2$ for Band 2 (950 to 1760 MHz), with 100 RMs/deg$^2$ and a nominal 10,000 deg$^2$ survey area  \citep{heald_magnetism_2020}. This FoM highlights the key scientific discovery advantage of an SKA-Low RM Grid, in particular for statistical studies of the low-density, weakly magnetised Universe. 

%However, for the study of extragalactic magnetic fields, it is not just the nominal measurement error that is important; an accurate removal of the foreground RM contribution from the Milky Way is also essential \citep[e.g.][]{cava25}. Presently, the estimated uncertainty propagated into the extragalactic RMs from the subtraction of the Galactic rotation measure (GRM) is of the order of 1 rad/m$^2$ at best. This is based on data dominated by the NVSS RM catalogue \citep{taylor_rotation_2009-1,van_eck_rmtable2023_2023}, which has a sky density of $\sim$1 RM/deg$^2$ and typical RM errors of $\sim$10 rad/m$^2$. 
However, for studies of extragalactic magnetic fields, it is not sufficient to consider only the nominal measurement uncertainty; an accurate subtraction of the Milky Way RM is also essential \citep[e.g.][]{cava25}. At present, the best Galactic rotation measure (GRM) models \citep{hutschenreuter_galactic_2021} propagate an uncertainty of order 1 rad/m$^2$ into the extragalactic RMs, even though they are primarily constrained by the NVSS RM catalogue \citep{taylor_rotation_2009-1,van_eck_rmtable2023_2023}, which has a low sky density ($\sim$1 RM/deg$^2$) and typical individual RM uncertainties of $\sim$10 rad/m$^2$. 
Considering that the SKA-Low RM Grid area would overlap with the POSSUM survey, one can expect at least an order of magnitude improvement in the accuarcy of the GRM estimation, especially on angular scales of less than 1 degree. This is because POSSUM has an RM number density 40 times that of the NVSS, in addition to almost 10 times smaller RM errors. The inclusion of the high-precision SKA-Low RMs, and SKA-Mid RMs where available, can further enhance the GRM models. A goal of having the residual GRM errors at least comparable to the SKA-Low RM measurement errors, in high Galactic latitude regions, may not be unrealistic. 
Improvements in algorithms for the statistical separation of the GRM and noise from the extragalactic RM will further improve the scientific inferences in this area \citep{Vacca02.2026.SKA}. 

%\textbf{Add a table showing FoM comparsions; like in POSSUM RASP doc}
%\textbf{Comment on GRM error, benefit of POSSUM (no equivalent northern hemisphere survey), and RM variance of extragalactic population being relative to properties of target (e.g. extragal bkg vs cosmic filament variances}. 

%\textbf{divide FoM by 1000 and comment on impact of GRM model error term for extragalactic studies, as well as importance of sigmaRM background.}
%\textbf{also consider that Piras counts are at 6$''$, more comparable to expected SKA-Low AA4 resolution}

\textbf{Survey strategy:} To maximise the RM Grid capability for extragalactic science, the host galaxies and redshifts are needed. Areas of sky with deep optical photometry and spectroscopy are therefore preferred for wide-area RM Grids. Considering the modest sensitivity difference at 5$''$ between AA* and AA4, the wide-area survey could begin with AA* and continue with AA4, potentially without repeating sky areas (assuming the data quality is good). The ideal extragalactic sky area would start with the DES area \citep[][$\sim$5,000 deg$^2$]{DES} and continue to expand outwards into the Euclid planned area \citep{euclid2022}. Overlap with the Simons Observatory Large Area Telescope survey \citep[][$\sim$16,000 deg$^2$]{SimonsObservatory} will also provide exciting synergies for characterising the magnetic fields properties of the ionized gas detected through the Sunyaev-Zeldovich effect (e.g.~high-$z$ galaxy clusters, dense phases of the Warm-Hot-Ionized Medium (WHIM), etc.). 
% Simons Obs Large Aperture Telescope (LAT): This 6-meter telescope is set to survey approximately 40% of the sky, equating to around 16,000 square degrees. It operates with arcminute resolution
The deep photometry of Euclid and the Rubin Telescope \citep{rubin2019} will further assist in identifying high redshift host galaxies. This has a significant scientific impact, as the strongest constraints on the models for the origin of cosmic magnetism and the strength of primordial magnetic fields come at the highest redshifts \citep{cava25}. 
%For example, only 14\% of the LoTSS DR2 RM Grid has sources with $z>1$ and $\lesssim1$\% at $z > 2$.

There is an opportunity to produce a high-impact early RM Grid using AA* Global Sky Model data\footnote{\url{https://www.skao.int/en/science-users/642/science-user-qa}} (e.g.~to a nominal depth of $\sim$30~$\mu$Jy/beam). We predict 2.6 RMs/deg$^2$ ($\sim$5 times LoTSS) for an $8\sigma_{QU}$ detection threshold, in regions outside the Galactic plane ($|b|>10^\circ$). The feasibility of this could be investigated during the Science Verification phase (2027--2029). Such an RM Grid would better inform the strategy for a deeper survey (both technically and scientifically), in relation to the optimal wide-area survey region and integration time that would provide the best scientific return. Given that SKA-Low can in principle observe $\sim$75\% of the sky, an ``all-sky'' survey combined with a deeper wide-area survey could eventually reach 100,000 RMs. 

\subsection{An SKA-Low ultra-deep RM Grid survey}
\label{sec:deep}
An SKA-Low AA4 deep survey would be close to confusion limited in Stokes $I$ after $\sim$30 hours (based on the calculator). However, given the low sky-density of polarized sources at m-wavelengths, it would not be confusion-limited in polarization \citep{heald_magnetism_2020}. Integrating on a single field for $\sim$1000 hours could in theory reach a noise level of $\lesssim$1~$\mu$Jy/beam in Stokes $Q$ and $U$. We do not predict the exact number of RMs based on Eqn.~\ref{eqn:pcounts} because this would extrapolate too far outside the known regime. However, it is clear that for a given survey time, a wide-area survey will provide a greater total number of RMs than an ultra-deep one. For example, even with a sky density of order 10 RMs/deg$^2$ and 4000 hours of integration time (on four fields, for 100 deg$^2$ coverage) the total number of RMs would remain an order of magnitude less than the wide-area survey. 
%\textbf{Sens calc gives 0.5 uJy).
%unfair for FoM comparison, need roughly same number of hours; e.g. 25 deg$^2$ for 10 hrs * 12 fields is 3,000 hrs. 
%1000 hrs per field for 3 fields makes more sense? 
%Do not use FoM, or predictions since this is too far beyond known properties? But do not focus much on deep field since more relevant for other science goals.}

%\textbf{Estimate number of pol sources at threshold of 8sigmaQU (or lower) for 10 uJy with 10k deg and 1 uJy with a few 100 deg area. What is the probability of detecting e.g. 100 sources at z > 3 for either survey?}

The scientific goals of an ultra-deep RM Grid are likely more related to the discovery potential for the types of polarized sources that would be detected, although targeted observations of individual structures (e.g.~a local cosmic web filament) could be rewarding \citep{vacca2024}. 
%\noindent For a detection threshold of 8 uJy, we predict $>20$ RMs per deg$^2$, which for example, implies over 1,500 RMs for three fields of 25 deg$^2$ each. 
%However, we consider this estimate highly uncertain as it extrapolates too far beyond the known polarized properties of sources at m-wavelengths. 
%While the total number of RMs and the FoM would not be competitive with the wide-area survey, the deep survey would provide exciting discovery potential. 
It would also provide a reliable path for future RM surveys by advancing our knowledge of the properties of the faintest polarized sources and their RMs \citep[e.g.][]{piras_lofar_2024}. It may be challenging to reliably identify the host radio galaxy of the RM due to Stokes $I$ confusion, which would be required to identify the optical host galaxy and redshift. Follow-up SKA-Low VLBI\footnote{\url{https://www.atnf.csiro.au/projects/instrumentation/lambda/}} observations could be necessary, and/or complementary SKA-Mid deep field observations \citep{loi2019}. 

Similar to an SKA-Low wide-area RM Grid, the full benefit of the signal-to-noise would not be accrued if the residual ionospheric RM errors remain at $\sim$0.05~rad/m$^2$. 
%In general, it is clear that the scientific power of SKA-Low RM surveys are significantly hindered by the residual ionosphere RM correction errors. 
Any improvements in the ionospheric RM modelling provided by, for example, TEC monitoring at the SKA-Low station locations would have a significant scientific impact.  
Minimising ionospheric contributions to the RM error budget improves RM accuracy and reduces depolarization from temporal ionospheric RM variability, thereby increasing both the recovered polarized flux and the number of detected polarized sources in RM grid surveys.
%Not having the ionosphere dominating the RM errors improves the RM accuracy, making such RM Grid surveys more scientifically rewarding, while also recovering more polarized flux and more sources by reducing the effect of depolarization. 
We note that a study of close pairs of RMs (Section~\ref{sec:DRMdz}) would not be subject to the ionospheric RM error when taking the difference of the RM pairs for the same observation.  

%\textbf{Comments on redshift distribution: I need to modify code to suppress high-z numbers to match current obs? Then remove high-z suppression for predictions?}
%The LoTSS DR2 RM Grid has 22 sources with z > 2. However, with deep optical imaging, such as provided by Euclid, LSST, etc.~these predictions may be severely underestimated (check by comparing with deep fields; do they see the steep fall-off like Martin in DR2 Stokes I?). 

%Given the more exploratory nature of an SKA-Low deep-field RM Grid, we do not consider it further for our science analysis below. 

%\begin{itemize}
%    \item EoR deep field might be 1000 hrs from 106 to 196 MHz; over what area? Details will appear in SKA-Low EoR data challenge paper. 
%\end{itemize}

\section{Updated constraints on the origin of cosmic magnetic fields}
\label{sec:newconstraints}

\begin{figure}
        \centering
        \includegraphics[width=0.99\linewidth,clip=true,trim=0.5cm 0cm 0cm 0cm]{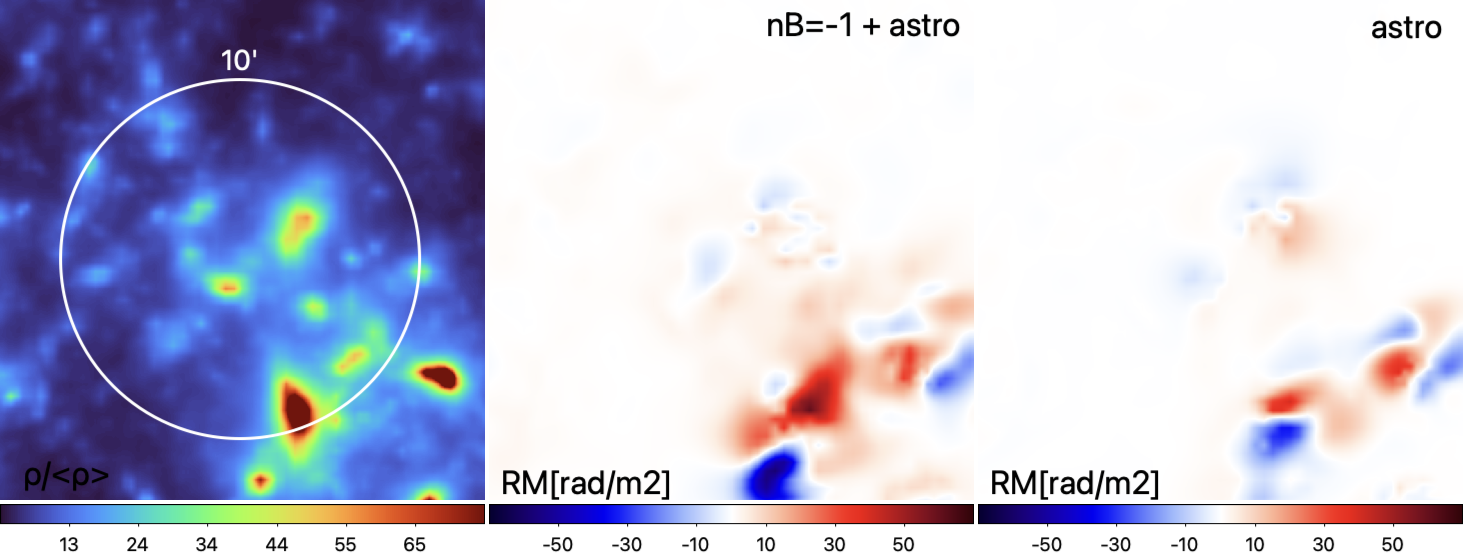}
        \includegraphics[width=0.99\linewidth,clip=true,trim=0.5cm 0cm 0cm 0cm]{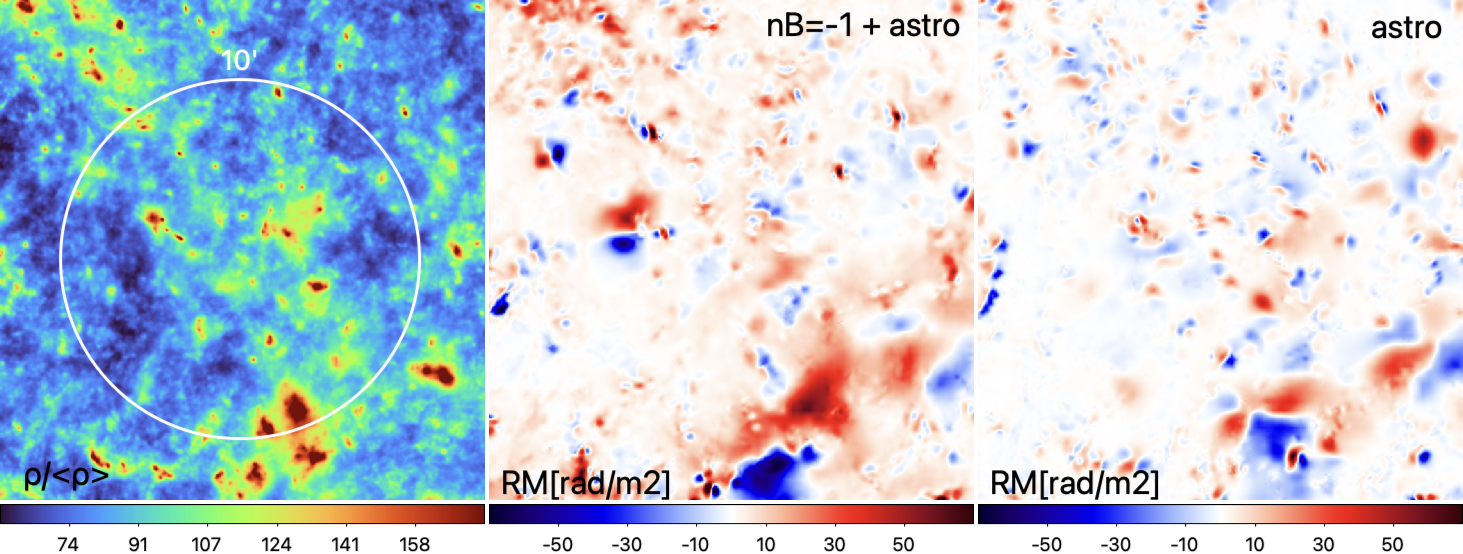}
        \caption{Left panels: projected average dark matter overdensity along the line of sight, for a simulated integration up to $z=0.1$ (top) or up to $z=2$ (lower panel). 
        Central and right panels: integrated RM up to $z=0.1$ or $z=2$, for our fiducial model comprising a primordial magnetic field model ($n_B=-1$ spectrum with $0.37 ~\rm nG$ normalisation, Section~\ref{sec:sim}) with an astrophysical seeding by galaxy processes (central panels), and for a comparison model only including astrophysical seeding of magnetic fields (right panels). 
        }
        \label{fig:RM_sim}
\end{figure}

\subsection{Overview of the cosmological MHD simulations used}
\label{sec:sim}
%\textbf{move closer to start of section 4? unify notation, i.e. $\rm \Delta RM^2$ vs $|\Delta{\rm RRM}(\Delta{z})|$ }\\
 To produce updated model predictions for comparison with the observed LOFAR extragalactic RM behaviour as a function of redshift (Sections~\ref{sec:RRMz},~\ref{sec:DRMdz}), we use a recent set of cosmological magneto-hydrodynamical (MHD) simulations, obtained with the cosmological Eulerian code ENZO\footnote{\url{https://enzo-project.org}}, and introduced in 
\citet[][]{va25a}. These simulations explore a broad suite of cosmological magnetization scenarios, including variations in both primordial and astrophysical seeding processes. The covered volume is $42^3\,\rm Mpc^3$ comoving, with a uniform  spatial resolution of $41.5$ kpc/cell. The simulations were  calibrated to reproduce several key observables of the galaxy distribution, which can be used to predict the maximal magnetization by feedback processes (from AGN and star formation) in a realistic way. The same simulations have been compared with LOFAR RM observations in \citet[][]{carretti_nature_2024}, hereafter C25, showing that no astrophysical magnetization mechanism, alone, can account for the observed RM behaviour at $z \geq 1$.

Here we use a single simulated magnetization scenario that best matches the current observational data \citep{cava25}, which we refer to as the baseline model. This scenario is predominantly primordial, but includes a realistic astrophysical contribution as a result of galaxy formation-related processes (Figure \ref{fig:RM_sim}). The stochastic primordial initialization of magnetic fields is obtained by drawing magnetic field vectors from a power-law spectrum: $P_B(k) dk \propto  k^{n_B} dk$, with $n_B=-1$ and with a normalization set to  $\sqrt{\langle B^2 \rangle}\approx 0.37~\mathrm{nG}$ after smoothing the field to a physical scale of $1 \rm ~Mpc$ (comoving), as usually done in CMB analysis  \citep[e.g.][]{2019JCAP...11..028P}. This particular normalization, set mainly by the comparison with the LOFAR RMs, yields a normalization $\sim 5$ times below existing CMB constraints. 
While the primordial magnetic field obtained in such a scenario dominates the magnetization in most of the cosmic volume, the addition of the injection of magnetic fields from star formation and AGN feedback, as explained in detail in \citet{va25a}, improves the comparison to the observed RM data at low redshift.

\subsection{RRM(z) analysis of the magnetised cosmic web}

\begin{figure}
        \centering
        \includegraphics[width=0.52\linewidth,clip=true,trim=0.5cm 0cm 0cm 0cm]{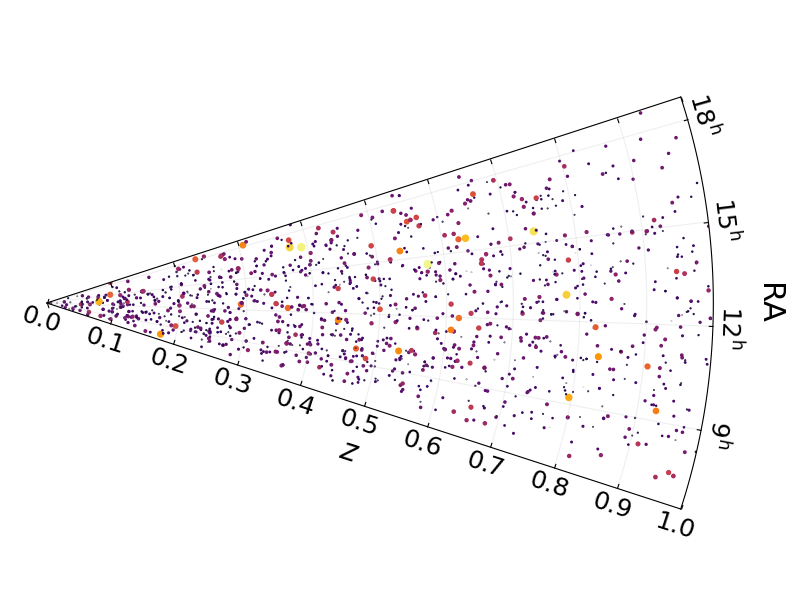}
        \includegraphics[width=0.47\linewidth,clip=true,trim=2.cm 0cm 0cm 0cm]{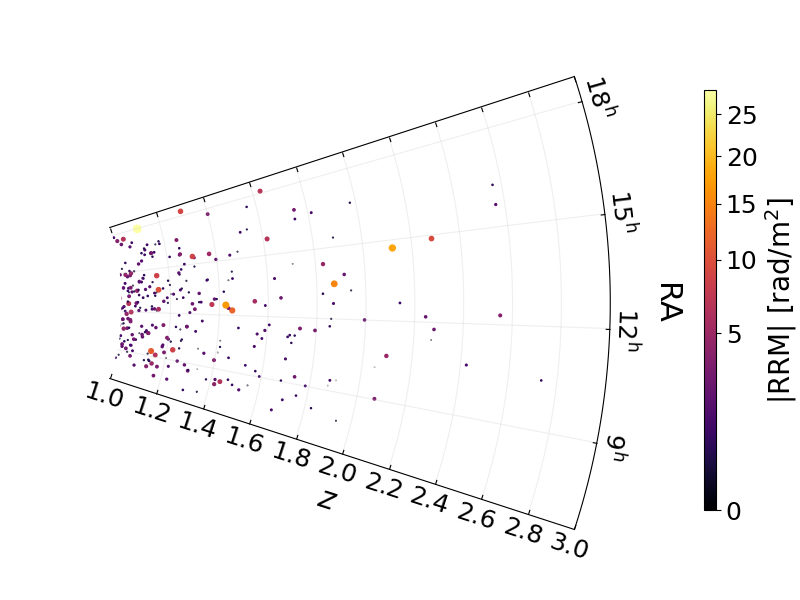}
        \caption{The RRM$(z)$ distribution from the LoTSS DR2 RM Grid, for $0<z<1$ (left) and $z>1$ (right). One can clearly see the sharp fall-off in the number of data points beyond $z=1$, which are crucial for constraining the origin of cosmic magnetism and the strength of primodial magnetic fields (Section~\ref{sec:newconstraints}). 
        }
        \label{fig:wedgeRRM}
\end{figure}

\label{sec:RRMz}
\subsubsection{Summary of previous results}
As shown in \citet{cava25}, and references therein, m-wavelength RM Grids are a powerful tool for the study of the origin of cosmic magnetism. 
They used the LoTSS DR2 RM Grid (OS23) to analyse the behaviour of the residual RM (RRM) as a function of redshift (Fig.~\ref{fig:wedgeRRM}) in comparison with expectations from cosmological MHD simulations of different magnetogenesis scenarios (Section~\ref{sec:sim}). Through this approach, they find a strong preference for the existence of sub-nG primordial seed magnetic fields. Key to this analysis is the separation of the GRM from the extragalactic RM. They used the `Faraday Sky' map of \citet{hutschenreuter_galactic_2021} to estimate the GRM (as the average map value within a 1 degree region surrounding each source) in order to subtract it from total RM (RRM $=$ RM $-$ GRM). 

They focused on a sub-sample of the LoTSS RMs, which had spectroscopic redshifts and |GRM|$<14$~rad/m$^2$. 
This was done in an attempt to minimise the propagated error from the GRM model into the RRM data. Future dense RM Grids from SKA Precursors, and the expected SKA-Mid RM Grid with up to 100 RMs/deg$^2$ \citep{Loi01.2026.SKA}, will be essential to robustly assess and isolate the GRM contribution on scales less than 1 degree. 
The final sub-sample contained 653 RRMs (with a median redshift of 0.47 and a maximum of 3.22), and displayed a correlation with redshift (linear fit slope of $0.25\pm0.08$), but with non-monotonic behaviour at $z <1$ (referred to as ``wiggles''). No obvious physical origin for these wiggles could be identified, and it remains to be determined if they persist in samples with higher statistical significance or not. These wiggles are not seen in the cosmological simulations but one physical possibility could be a signature of supercluster structures \citep[e.g.][]{pignataro_detection_2025,alonsolopez2026}, which can only be sampled in cosmological volumes much larger than provided by most MHD simulations. 

The total rms RRM of this sub-sample is $1.54\pm0.05$~rad/m$^2$, which is substantially lower than the typical extragalactic rms RRM of 6 to 8~rad/m$^2$ found from cm-wavelength RM Grids \citep[][]{schnitzeler_latitude_2010-1,oppermann_estimating_2015-1,anderson2024}. This is generally interpreted as a selection effect on the m-wavelength RMs caused by Faraday depolarization, which strongly suppresses any polarised emission with RM variations $>0.3$~rad/m$^2$ within the telescope angular resolution \citep[][]{stuardi_lofar_2020,mahatma_low_2020}. While this makes the detection of large samples of polarized sources at m-wavelengths very challenging, the upside is that those that are detected typically probe more underdense lines-of-sight (LoS) through the Universe than at cm-wavelengths, which is ideal for studies of cosmic web filaments. 
This interpretation is supported by the identification of a sub-sample of LoTSS RMs whose LoS pass within the virial radius of known galaxy clusters and have an rms RRM of $\sim$7~rad/m$^2$ (consistent with the extragalactic background from cm-wavelength RM Grids). These sources contribute only $\sim$10\% to the total rms RRM. Further detailed comparisons with known intervening structures leads \citet{carretti_nature_2024} to estimate contributions up to 20\% from clusters, groups and intervening (bright) galaxies to the total rms RRM. 
% sqrt(1.5^2-0.7^2) = 1.3 rad/m$^2$
Independent RM pairs studies (Section~\ref{sec:DRMdz}) indicate that only $\sim$15\% of the total rms is produced in the local environment of the radio galaxies \citep{pomakov_redshift_2022-1}. 
These studies thus imply that the majority of the rms RRM is most likely due to the magnetised WHIM of cosmic web filaments. 

\subsubsection{Updated results}
To quantify the RM contribution of cosmic web filaments, C25 use a model of the filament RM contribution (RRM$_f$) in addition to the A$_{rrm}$ parameter designed to account for other astrophysical components constant with redshift, either local or by intervening objects. While C25 combined these terms linearly, we employ a slight modification here by combining them in quadrature as 
\begin{equation}\label{eqn:RRMz}
   {\rm RRM_{rms}} =  \langle {\rm RRM}^2 \rangle^{1/2} = \sqrt{\left(\frac{A_{rrm}}{(1+z)^2}\right)^2 + RRM_f^2 }
\end{equation}
with
\begin{equation}
   RRM_f = \int_z^0 \frac{n_e(z^\prime)B_{||,f}(z^\prime)}{(1+z^\prime)^2}dl,
\end{equation}
where $n_e(z)$ is taken directly from the density distribution of the cosmological simulations and the physical magnetic field strength of each filament is defined as $B_f(z) = B_{0,f}(1+z)^\alpha$. 
The simulation LoS are restricted to matter overdensities that best match our estimate of the overdensities probed by the LoTSS RM observations (see C25 for details). 
%Note that a virial radius of $r_{100}$ corresponds to $\delta_M\sim160$. 
Therefore, the model only contains three free parameters: the magnetic field strength in filaments at redshift $z=0$, the power-law index $\alpha$ and the A$_{rrm}$ term. 
%\textcolor{blue}{TBD: updated results from Ettore, or summary of Carretti\&Vazza.}
Repeating the analysis for 100 $n_e(z)$ realisations from the same simulation\footnote{The differences in $n_e(z)$ between simulations are minimal, as shown in \citet{carretti2023}.}, we find $B_{0,f}=40\pm8$ nG and $\alpha=1.0\pm0.6$. This implies, for example, a comoving filament field strength of $B_{c,f}\sim27$~nG at the median redshift $z_{\rm med}=0.47$. The best-fit A$_{rrm}$ is $1.24\pm0.11$~rad/m$^2$, which indicates $\sim$15\% of non-filament contributions to the total RRM rms, again for $z_{\rm med}=0.47$.

\subsection{$\Delta$RRM$(\Delta{z})$ analysis of the magnetised cosmic web}
\label{sec:DRMdz}
One of the most challenging aspects of the RRM$(z)$ analysis (Section~\ref{sec:RRMz}) is the robust isolation of the GRM from the true extragalactic RM. 
An alternative approach that minimises the impact of the GRM, is to take the RM difference between close pairs of sources (or source components) on the sky. In regions away from the Galactic plane ($|b|>20^o$), the contribution of the GRM to the difference between RMs on small angular scales ($\theta<10'$) is expected to be small or sub-dominant compared to the extragalactic RM \citep[e.g.][]{leahy_small-scale_1987-1,stil2011}.  
% Leahy study was for scales < 3 arcmin! 
\subsubsection{Summary of previous results}
This approach was used in \citet{vernstrom_differences_2019-3}, using RM observations at 1.4 GHz, to analyse extragalactic magnetic fields by splitting their RM sample into physical pairs (i.e.~double-lobed radio galaxies) and non-physical or ``random'' pairs (i.e.~sources at different redshifts but close in projection on the sky). 
They found a large and significant difference between the RM distributions of the physical pairs (PP) and the non-physical pairs (NPP). In particular, for overlapping angular scales (3 to 11 arcmin), the NPP rms RM was larger by $\sim$5~rad/m$^2$, which can be most straightforwardly interpreted as the RM contribution from the intergalactic medium (IGM) along the additional path lengths probed by the NPP compared to the PP. 
Inferring the IGM magnetic field properties from this measurement is challenging without accounting for realistic distributions of cosmic structures along the different lines of sight probed by these observations. 
%\textbf{what overdensity cut would correspond to the 1.4 GHz observations? can we say if this is dominated by galaxy groups/clusters or not?}

Similar RM pair studies were conducted by \citet{osullivan_new_2020} and \citet{pomakov_redshift_2022-1} but at 150 MHz using the LoTSS DR2 RM Grid. They found a smaller RM difference between the PP and NPP samples of $\sim$1.6~rad/m$^2$. The difference between the m-wavelength and cm-wavelength results can be attributed to the systematically different lines of sight (LoS) probed by the respective RMs. For example, sources that are embedded in rich environments (with large local variations in magnetic field and gas density) or whose LoS pass through such environments, can experience small-scale RM variations of 10s of rad/m$^2$, which will produce significant Faraday depolarization.  While this reduces the observed degree of polarization at cm-wavelengths, it can suppress the polarized emission at m-wavelengths enough to make these sources effectively undetectable. Therefore, the m-wavelength RMs are missing the contributions from overdense structures such as groups and clusters of galaxies, making them more precise probes of the unbound gas in cosmic web filaments. 

\subsubsection{Updated observational results}
\begin{figure}
        \centering
        \includegraphics[width=0.48\linewidth,clip=true,trim=0.3cm 0cm 0cm 0cm]{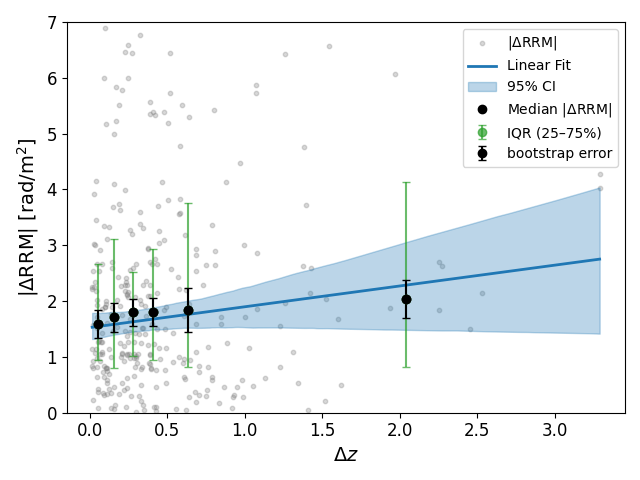}
        \includegraphics[width=0.49\linewidth,clip=true,trim=0.5cm 0cm 0cm 0cm]{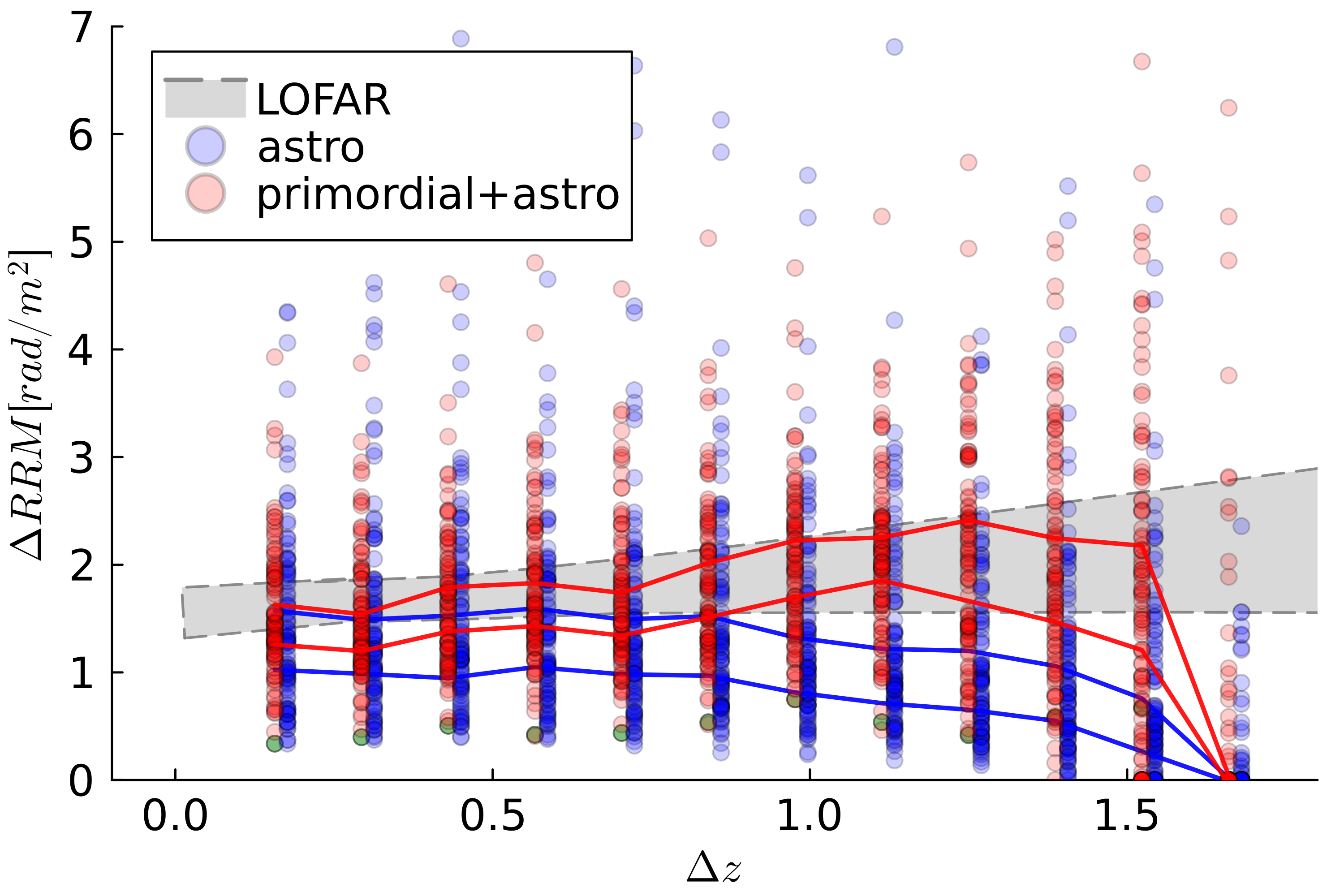}
        \caption{Left: Observational data from the LoTSS DR2 RM Grid of $|\Delta$RRM$(\Delta{z})|$ of non-physical pairs. A linear fit to the data gives a slope of $\sim$0.3, however the data remain consistent with a flat behaviour. 
        The median $|\Delta$RM$(\Delta{z})|$ are shown for six bins with equal numbers of sources per bin (57), with their bootstrap errors and the interquartile range (IQR). 
        Right: Simulated $|\Delta$RRM$(\Delta{z})|$ for the baseline (primordial+astrophysical) magnetogensis model (red), and for a model using only astrophysically seeded magnetic fields (blue). The solid lines for each model give the range of the bootstrap errors obtained as in the left image, while the colored points give the distribution of generated random pairs. The grey area show the existing constraints from LOFAR (as in the left plot).
        }
        \label{fig:deltaRM_zeta}
\end{figure}

Based on the same LOFAR data, we show in Figure~\ref{fig:deltaRM_zeta}~(left) the median $|\Delta{\rm RRM}|$ for the NPP as a function of redshift difference ($\Delta{z}$) with an equal number of sources per bin (57). We show the bootstrap errors on the medians (to compare with the median values from the simulations), as well as the interquartile range (IQR) of the $|\Delta{\rm RRM}|$ data per bin (to compare with the intrinsic scatter of the simulation data). There is a hint of a correlation in the $|\Delta{\rm RRM}(\Delta{z})|$ data, with a linear fit yielding a slope of $\sim$0.3. However, the slope is consistent with 0 within the 95\% confidence interval. 

\subsubsection{New simulation of $\Delta\mathrm{RRM}(\Delta z)$ for SKA-like Observations}
\label{sec:newtests}
To investigate constraints that could be provided by the SKA telescope capabilities, we simulated the observable properties of the $\Delta {\rm RRM}(\Delta z)$ statistics, using the simulations described in Section~\ref{sec:sim}. This was done by generating a series of light cones with the integrated RM along the line of sight (LoS), and by placing two sources at randomly different redshift differences along the LoS, in order to mimic the effect of the Faraday Rotation by intergalactic magnetic fields, similar to previous work \citep[][]{osullivan_new_2020,va21}. Each generated light cone has an aperture of $20'$, a angular resolution of $\approx 2"$, and it integrates the simulated comoving volume up to $z=2$, by opportunely rescaling the comoving simulated volume at increasing redshift based on the angular scale distance for the assumed cosmology, and after applying the redshift-dependent correction factors. We removed from every LoS the contribution from cells in
the simulation with a gas matter over-density larger than $50$ times
the cosmic average gas density, to minimise the RM contamination by cluster halos and 
approximately match the estimated density range of LoTSS RM observations 
\citep{carretti2023}. 
%(see Section~\ref{sec:RRMz} ).
Finally, at every crossing of the comoving volume, the field of view is randomly shifted in the image plane to avoid the production of artifacts connected to the repetition of the same cosmic structures along the LoS \citep[see e.g.][]{2024ApJ...977..128M}.

Figure \ref{fig:RM_sim} shows the RM along the line of sight in our simulation, integrating the volume out to $z=0.1$ (top row) or out to $z=2$ (bottom row). For comparison, we also show the integrated dark matter density in the same volume, and the RM is calculated for our aforementioned baseline model (including both a primordial and an astrophysical seeding of magnetic fields, Section~\ref{sec:sim}) and a comparison run in which only the astrophysical seeding is included.

\subsubsection{Comparison of simulated $\Delta\mathrm{RRM}(\Delta z)$ with observational data}
\label{sec:simobs}

Figure~\ref{fig:deltaRM_zeta}~(right) shows the simulated relation of the median $|\Delta{\rm RRM}|$ between non-physical pairs as a function of their redshift difference ($\Delta z$), for the baseline magnetogenesis model, and for a model only using astrophysical seeding. Both are contrasted with existing constraints from LOFAR \citep{pomakov_redshift_2022-1} and Figure~\ref{fig:deltaRM_zeta}~(left; grey shaded region).  The angular separation of the two sources is randomly chosen in the $4'' \leq \theta \leq 10'$ range, and their two redshifts are also independently drawn in the $0.02 \leq z \leq 2$ range. 
The median of both scenarios is not far from LOFAR observations, 
consistent with our previous work \citep[][]{osullivan_new_2020}, but the data show a statistical preference for the  baseline primordial model, which within the errors is consistent with the LOFAR data. 

We also report a significantly larger scatter in the astrophysical scenario, reflecting the smaller filling factor of magnetic fields in the cosmic volume in this latter case, where magnetic ``bubbles'' fill less than $\sim 20 \%$ of the volume \citep[][]{va25a}. 
%This provides another means to constrain the models with future larger samples of observational data. 
We also report a significant drop of the relation at large $\Delta z$, which is explained by the fact that the peak of magnetisation from stars and AGN is in the $z \sim 2$ range:  so the absolute value of $|\Delta \rm RRM|$ tends to be smaller for large $\Delta z$, as one of the two sources is far from the maximum of magnetisation by purely astrophysical sources in our model (considering that our simulated volume only reaches $z=2$).  
This trend, as well as the larger scatter, represent additional tools to further disentangle primordial from astrophysical models with the order-of-magnitude more RM data provided by SKA-Low.  
Figure~\ref{fig:deltaRM_zeta} shows the significant scatter in both observed and simulated data, and we estimate that $\sim$4 to 9 times more sources are needed (if errors scale as $\sqrt{N}$) to robustly discriminate between the two simulation scenarios. 

\section{Advances provided by an SKA-Low wide-area RM Grid}
\label{sec:advances}

The expected RM number density of $\sim$5.9 RMs/deg$^2$ from an SKA-Low wide-area RM Grid (Section~\ref{sec:polcounts}) provides an exciting opportunity to significantly advance our knowledge beyond the current best m-wavelength RM Grids ($\sim$0.4 RMs/deg$^2$). In terms of the study of extragalactic magnetic fields, one essential aspect is to identify the host radio sources of the RM as well as obtaining the ancillary host galaxy properties of the radio sources. 

As the polarized sources are part of the relatively bright radio galaxy population, the expected excellent image fidelity and good angular resolution of SKA-Low should make the identification (ID) of the host galaxy more reliable. Several automated and semi-automated procedures for obtaining host IDs have been developed \citep[e.g.][]{hardcastle_lofar_2023,gupta_rg-cat_2024,gordon_quick_2023} and continue to be developed further. 
Additionally, given the current and planned deep and wide-area optical imaging and spectroscopic surveys of the southern sky (e.g.~DES, DESI, 4MOST: \cite{4most2025}, Euclid, Rubin Telescope), we can expect to match or better the host galaxy ID and redshift statistics of the LoTSS RM Grid (89\% with host IDs, 80\% with redshift estimates). This would mean at least 45,000 RMs with redshift estimates; more than 20 times greater than in the LoTSS DR2 RM Grid. The deep optical surveys should also enhance the $z>1$ statistics, meaning a potentially larger relative gain is expected at high redshifts, where the strongest constraints on magnetogenesis scenarios can be obtained. 

These greater statistics open up an array of additional options to analyse the RM($z$) data, and will allow us to control for many possible systematic effects (e.g.~source type/morphology, sky position, GRM quality, etc.). In turn, this enables more robust inferences of the properties of magnetic fields in cosmic web filaments and thus the nature of primordial magnetic fields, in addition to better constraining the RM properties of intervening objects (e.g.~galaxy halos, intercluster and intergroup media). 
For example, with many more RMs one could conduct a full Bayesian fit for all free parameters in Eqn.~\ref{eqn:RRMz} to constrain both the astrophysical RM evolution as well as that from cosmic web filaments (Section~\ref{sec:RRMz}). 
We could determine the RRM$(z)$ trend with much more precision, with lower statistical errors in each bin. This would allow us to confirm the existence of the ``wiggles'' and to determine their nature, to search for new features, and to test for any curvature (i.e.~currently we can only constrain a linear fit). Furthermore, the complementary $\Delta{\rm RRM}(\Delta{z})$ analyses can provide independent tests of the magnetogenesis scenarios (Section~\ref{sec:DRMdz}). 

%\textbf{How does a 15x improvement in RM numbers help!? High z stats, go from 2sigma to 9sigma constraint on 0.1 nG? Do wiggles remain at low-z, do they continue to high-z? Measure extent of astro-feedback/magnetised bubbles (how: split LoS as a function of cosmic overdensity)? Help distinguish between feedback models of different simulations; veering into SKA-Mid territory. }

\textbf{Cross-correlation with filament catalogues:}
Statistical detections of cosmic web filaments have been found using X-ray, the thermal Sunyaev-Zeldovich effect \citep{tanimura_first_2020,tanimura_density_2019} and radio synchrotron emission \citep{vernstrom_discovery_2021} by cross-correlation with optical galaxies as filament tracers. RM cross-correlations have only been able to provide upper-limits thus far \citep{amaral_constraints_2021}, due to the limited quality and quantity of the RM data. Considering the exceptional advances in the identification and study of cosmic web filaments with, for example Euclid, significant detections should be expected by cross-correlating with an SKA-Low RM Grid in the overlapping sky area. 
Cross-correlations with local Universe structures will also be essential to identify the contributions of, for example, nearby superclusters of galaxies \citep{pignataro_detection_2025,alonsolopez2026}. 

\textbf{Magnetic field of individual filaments:}
There are several recent X-ray detections of individual filaments  between galaxy clusters \citep{dietl_discovery_2024,veronicaEROSITAViewAbell2024,migkasDetectionPureWarmhot2025a}, with properties closer to those expected from the WHIM, as opposed to denser regions such as intercluster bridges \citep{govoni2019,botteon2020,venturi_radio_2022,pignataro_detection_2025,Gustafsson2026}. Some of these nearby filaments have a large enough angular size (several deg$^2$) such that their expected RM dispersion of 0.5 to 1 rad/m$^2$ \citep{carretti_magnetic_2022}, could be detectable as long as a sufficient number of RMs can cover the filament 
(e.g.~30 RMs would give a statistical precision of $\sim0.3$~rad/m$^2$ on the m-$\lambda$ extragalactic background RM of $\sim$1.5 rad/m$^2$). 
%(e.g.~30 RMs would enable a 3.6$\sigma$ detection of an excess filament dispersion of 1 rad/m$^2$ over the m-$\lambda$ extragalactic background RM of $\sim$1.5 rad/m$^2$. 
However, this would likely require a negligible GRM error contribution, which may prove challenging. 
%provide very small statistical errors of $\delta{\rm RM}/\sqrt{30} \simeq 0.01$~rad/m$^2$).  

\textbf{Broader applications:} 
Of course m-$\lambda$ RM Grids can be, and have been, used for many other science goals, not just determining the origin of cosmic magnetism as we focus on here. 
For example, high-precision RMs contribute to the study of the Milky Way magnetic field structure \citep[e.g.][]{hutschenreuter_galactic_2021,ungerfarrar2024}, while they can also be used to determine the maximum spatial extent of AGN feedback \citep{blunier_constraint_2024}, the properties of the magnetised circumgalactic medium in nearby galaxies \citep{heesenDetectionMagneticFields2023a}, and the physics of radio AGN and their environments \citep{mahatma_low_2020,stuardi_lofar_2020}. They also provide a means to discover new pulsars and transients \citep{sobey2022}, and may provide useful constraints on the possible existence and nature of dynamical dark energy \citep{naokawa_universal_2025}. 

\section{Conclusion}
\label{sec:conclusion}
%An SKA-Low RM Grid can enable an dramatic step-change in our understanding of the magnetised Universe by providing 10 to 20 times more high-precision RMs than the current best metre-wavelength surveys. 
An SKA-Low RM Grid can enable a dramatic step-change in our understanding of the magnetised Universe by providing possibly up to 100,000 high-precision RMs over the SKA-Low-observable sky ($\sim$20 times more than the current best metre-wavelength surveys). 
Combined with the upcoming deep optical photometric surveys and new spectroscopic data across the southern sky, this RM Grid will enable a wide range of investigations, from pulsars to early-Universe physics. In this chapter we have focused on the robust determination of the magnetic field evolution in cosmic web filaments. We find that an SKA-Low RM Grid will help solve some long-standing open problems in extragalactic astrophysics, such as the extent of magnetic pollution of the cosmic web by galaxies, and the origin of cosmic magnetism. 

\section*{Acknowledgements}
SPO acknowledges support from the Comunidad de Madrid Atracción de Talento program via grant 2022-T1/TIC-23797, and grant PID2023-146372OB-I00 funded by MICIU/AEI/10.13039/ 501100011033 and by ERDF, EU.  FV acknowledges funding under the European Union’s Horizon Europe program through the ERC Synergy Grant COSMOMAG (Project Id.~101224803), and the CINECA award  "IscrB\_CREW"  under the ISCRA initiative, for the availability of high-performance computing resources and support. 
VV acknowledges support from the Prize for Young Researchers ``Gianni Tofani'' second edition, promoted by INAF-Osservatorio Astrofisico di Arcetri (DD n.~84/2023).

\bibliographystyle{abbrvnat-maxbibnames4}
%\bibliography{franco}
%\bibliography{../enriched,../igmf,franco} 

\end{document}